\documentclass[12pt,a4paper]{article}
\makeatletter
\def\eqnarray{%
\stepcounter{equation}%
\let\@currentlabel=\theequation
\global\@eqnswtrue
\global\@eqcnt\z@
\tabskip\@centering
\let\\=\@eqncr
$$\halign to \displaywidth\bgroup\@eqnsel\hskip\@centering
$\displaystyle\tabskip\z@{##}$&\global\@eqcnt\@ne
\hfil$\displaystyle{{}##{}}$\hfil
&\global\@eqcnt\tw@$\displaystyle\tabskip\z@{##}$\hfil
\tabskip\@centering&\llap{##}\tabskip\z@\cr}
\makeatother
\newcommand{\ket}[1]{{\vert{#1}\rangle}}
\newcommand{\bra}[1]{{\langle{#1}\vert}}

\newcommand{\calh}{{\cal H}}

\newcommand{\fukuso}{{\mathbf C}}
\newcommand{\real}{{\mathbf R}}
\newcommand{\futon}{{\bf N}}

\newcommand{\zetta}{{\vert z\vert}}

\newcommand{\kappazetta}{{\vert\kappa\vert}}

\begin{document}

\title{\sl Matrix Elements of Generalized Coherent Operators}
\author{
  Kazuyuki FUJII
  \thanks{E-mail address : fujii@yokohama-cu.ac.jp }\\
  Department of Mathematical Sciences\\
  Yokohama City University\\
  Yokohama, 236-0027\\
  Japan
  }
\date{}
\maketitle\thispagestyle{empty}
%
%
%
%
\begin{abstract}
  Explicit forms are given of matrix elements of generalized coherent 
  operators based on Lie algebras su(1,1) and su(2). We also give a kind of 
  factorization formula of the associated Laguerre polynomials. 
\end{abstract}

\newpage

%
%
%
%

\section{Introduction}

Coherent states or generalized coherent states play an important role in 
quantum physics, in particular, quantum optics, see \cite{KS} and 
references therein, or the books \cite{MW}, \cite{WPS}. 
They also play an important one in mathematical physics. See 
the textbook \cite{AP}. For example, they are very useful in performing 
stationary phase approximations to path integral, \cite{FKSF1}, 
\cite{FKSF2}, \cite{FKS}. 

Coherent operators which produce coherent states are very useful because 
they are unitary and easy to handle. Why are they so handy ? 
The basic reason is probably that they are subject to the elementary 
Baker-Campbell-Hausdorff formula. 
Many basic properties of them have been investigated, see 
\cite{MW}, \cite{WPS}, \cite{AP} or \cite{KF6}, \cite{KF7}. 
We are particularly interested in the following three 
ones : the matrix elements, the trace formula and the Glauber's formula. 

Generalized coherent operators which produce generalized coherent states 
are also useful. But the corresponding properties have not been 
investigated as far as the author knows, see for example \cite{KF5}. 
One of the reasons is that they are not easy to handle. 
Of course we have disentangling formulas corresponding to the elementary 
Baker-Campbell-Hausdorff formula, but they are still not handy.

In this paper we focus our attention only on matrix elements of generalized 
coherent operators based on Lie algebras $su(1,1)$ and $su(2)$, and 
give explicit forms to all of them. 

We make a short comment on some applications. Matrix elements of coherent 
operator can be applied to Rabi oscillations in Quantum Optics, see 
\cite{MSIII} and \cite{MFr}. One of our aims is also to apply the calculations 
in this paper to the same subject. This will be published in \cite{KF15}.

\section{Coherent and Generalized Coherent Operators}

We review the general theory of both the coherent operator and 
generalized coherent ones based on Lie algebras $su(1,1)$ and $su(2)$.

\subsection{Coherent Operator}
Let $a(a^\dagger)$ be the annihilation (creation) operator of the harmonic 
oscillator.
If we set $N\equiv a^\dagger a$ (:\ number operator), then
\begin{equation}
  \label{eq:2-1-1}
  [N,a^\dagger]=a^\dagger\ ,\
  [N,a]=-a\ ,\
  [a^\dagger, a]=-\mathbf{1}\ .
\end{equation}
Let $\calh$ be a Fock space generated by $a$ and $a^\dagger$, and
$\{\ket{n}\vert\  n\in\futon\cup\{0\}\}$ be its basis.
The actions of $a$ and $a^\dagger$ on $\calh$ are given by
\begin{equation}
  \label{eq:2-1-2}
  a\ket{n} = \sqrt{n}\ket{n-1}\ ,\
  a^{\dagger}\ket{n} = \sqrt{n+1}\ket{n+1}\ ,
  N\ket{n} = n\ket{n}
\end{equation}
where $\ket{0}$ is the normalized vacuum ($a\ket{0}=0\  {\rm and}\  
\langle{0}\vert{0}\rangle = 1$). From (\ref{eq:2-1-2}) the state 
$\ket{n}$ for $n \geq 1$ is given by
\begin{equation}
  \label{eq:2-1-3}
  \ket{n} = \frac{(a^{\dagger})^{n}}{\sqrt{n!}}\ket{0}\ .
\end{equation}
These states satisfy the orthonormality and completeness conditions
\begin{equation}
  \label{eq:2-1-4}
   \langle{m}\vert{n}\rangle = \delta_{mn}\ ,\quad \sum_{n=0}^{\infty}
   \ket{n}\bra{n} = \mathbf{1}\ . 
\end{equation}

\noindent{\bfseries Definition}\quad We call a state  
\begin{equation}
\label{eq:2-1-7}
\ket{z} =  \mbox{e}^{za^{\dagger}- \bar{z}a}\ket{0}\equiv U(z)\ket{0} 
\quad \mbox{for}\quad z\ \in\ \fukuso 
\end{equation}
the coherent state.

Since the operator 
$
U(z) = \mbox{e}^{za^{\dagger}- \bar{z}a}
$ 
is unitary, we call this a (unitary) coherent operator. For these 
operators the following property is crucial : 
\begin{equation}  
  \label{eq:2-1-9}
  U(z+w) = \mbox{e}^{-\frac{1}{2}(z\bar{w}-\bar{z}w)}\ U(z)U(w) 
  \quad \mbox{for} \quad z,\ w \in \fukuso .
\end{equation}
From this we have a well--known commutation relation 
$
U(z)U(w) = \mbox{e}^{z\bar{w}-\bar{z}w}\ U(w)U(z)
$. 
Here let us recall the disentangling formula of $U(z)$ for later 
convenience : 
\begin{equation}  
  \label{eq:2-1-11} 
\mbox{e}^{za^{\dagger}- \bar{z}a}
=\mbox{e}^{-\frac{1}{2}{\vert{z}\vert}^{2}}\mbox{e}^{za^{\dagger}}
                                           \mbox{e}^{-{\bar z}a}
=\mbox{e}^{\frac{1}{2}{\vert{z}\vert}^{2}}\mbox{e}^{-{\bar z}a}
                                          \mbox{e}^{za^{\dagger}}.
\end{equation}
This is obtained by the famous (elementary) Baker--Campbell--Hausdorff 
formula : 
\begin{equation}
  \label{eq:2-1-12}
 \mbox{e}^{A+B}=\mbox{e}^{-\frac1{2}[A,B]}\mbox{e}^{A}\mbox{e}^{B}
\end{equation}
whenever $[A,[A,B]] = [B,[A,B]] = 0$.

\subsection{Generalized Coherent Operator Based on $su(1,1)$}
Let us introduce generalized coherent operators and the states based on 
$su(1,1)$. 
Let $\{ k_{+}, k_{-}, k_{3} \}$ be a Weyl basis of Lie algebra $su(1,1) 
\subset sl(2,\fukuso)$, 
\begin{equation}
  \label{eq:2-2-1}
 k_{+} = \left(
        \begin{array}{cc}
               0 & 1 \\
               0 & 0 \\
        \end{array}
       \right),
 \quad 
 k_{-} = \left(
        \begin{array}{cc}
               0 & 0 \\
              -1 & 0 \\
        \end{array}
       \right),
 \quad 
 k_{3} = \frac12 
       \left(
        \begin{array}{cc}
               1 &  0  \\
               0 & -1 \\
        \end{array}
       \right). 
\end{equation}
Then we have 
\begin{equation}
  \label{eq:2-2-2}
 [k_{3}, k_{+}]=k_{+}, \quad [k_{3}, k_{-}]=-k_{-}, 
 \quad [k_{+}, k_{-}]=-2k_{3}.
\end{equation}
We note that $(k_{+})^{\dagger}=-k_{-}$.

Next we consider a spin $K\ (> 0)$ representation of $su(1,1) 
\subset sl(2,\fukuso)$ and set its generators 
$\{ K_{+}, K_{-}, K_{3} \}\ ((K_{+})^{\dagger} = K_{-})$, 
\begin{equation}
  \label{eq:2-2-3}
 [K_{3}, K_{+}]=K_{+}, \quad [K_{3}, K_{-}]=-K_{-}, 
 \quad [K_{+}, K_{-}]=-2K_{3}.
\end{equation}
We note that this (unitary) representation is necessarily infinite dimensional.
 The Fock space on which $\{ K_{+}, K_{-}, K_{3} \}$ act is 
$\calh_K \equiv \{\ket{K,n} \vert n\in\futon\cup\{0\} \}$ and 
whose actions are
\begin{eqnarray}
  \label{eq:2-2-4}
 K_{+} \ket{K,n} &=& \sqrt{(n+1)(2K+n)}\ket{K,n+1},\ 
 K_{-} \ket{K,n}  = \sqrt{n(2K+n-1)}\ket{K,n-1} ,  \nonumber  \\
 K_{3} \ket{K,n} &=& (K+n)\ket{K,n}, 
\end{eqnarray}
where $\ket{K,0}$ is a normalized vacuum ($K_{-}\ket{K,0}=0$ and 
$\langle K,0|K,0 \rangle =1$). We have written $\ket{K,0}$ instead 
of $\ket{0}$  to emphasize the spin $K$ representation, see \cite{FKSF1}. 
From (\ref{eq:2-2-4}), states $\ket{K,n}$ are given by 
\begin{equation}
  \label{eq:2-2-5}
 \ket{K,n} =\frac{(K_{+})^n}{\sqrt{n!(2K)_n}}\ket{K,0} ,
\end{equation}
where $(a)_n$ is the Pochhammer's notation 
$
  \label{eq:2-2-6}
 (a)_n \equiv  a(a+1) \cdots (a+n-1) 
$. 
These states satisfy the orthonormality and completeness conditions 
\begin{equation}
  \label{eq:2-2-7}
  \langle K,m \vert K,n \rangle =\delta_{mn}, 
 \quad \sum_{n=0}^{\infty}\ket{K,n}\bra{K,n}\ = \mathbf{1}_K.
\end{equation}
Now let us consider a generalized version of coherent states : 

\noindent{\bfseries Definition}\quad We call a state  
\begin{equation}
   \label{eq:2-2-8}
 \ket{z} = \mbox{e}^{zK_{+} - \bar{z}K_{-}}\ket{K,0} 
    \equiv V(z)\ket{K,0}
  \quad \mbox{for} \quad z \in \fukuso.
\end{equation}
the generalized coherent state (or the coherent state of Perelomov's 
type based on $su(1,1)$ in our terminology).

\par \noindent  
Here is the disentangling formula :
\begin{equation}
  \label{eq:2-2-11}
    \mbox{e}^{zK_{+} -\bar{z}K_{-}}  
  = \mbox{e}^{\zeta K_{+}}\mbox{e}^{\log (1-\vert\zeta\vert^2)K_{3}}
    \mbox{e}^{-\bar{\zeta}K_{-}}
  = \mbox{e}^{-\bar{\zeta}K_{-}}\mbox{e}^{-\log (1-\vert\zeta\vert^2)K_{3}}
    \mbox{e}^{\zeta K_{+}},\quad 
    \zeta \equiv \frac{\mbox{tanh}(\zetta)}{\zetta}z.
\end{equation}
This is also the key formula for generalized coherent operators.  
See \cite{AP} or \cite{FS}.

Here let us construct an important example of this representation. 
If we set 
\begin{equation}
  \label{eq:2-2-12}
  K_{+}\equiv{1\over2}\left( a^{\dagger}\right)^2\ ,\
  K_{-}\equiv{1\over2}a^2\ ,\
  K_{3}\equiv{1\over2}\left( a^{\dagger}a+{1\over2}\right)\ ,
\end{equation}
then it is easy to check (\ref{eq:2-2-3}). 
That is, the set $\{K_{+},K_{-},K_{3}\}$ gives unitary representations of 
$su(1,1)$with spin $K = 1/4\ \mbox{and}\ 3/4$, \cite{AP}. 
Now we in particular call an operator 
\begin{equation}
  \label{eq:2-2-14}
   S(z) = \mbox{e}^{\frac{1}{2}\{z(a^{\dagger})^2 - \bar{z}a^2\}}
   \quad \mbox{for} \quad z \in \fukuso 
\end{equation}
the squeezed operator. This operator plays a very important role.

\subsection{Generalized Coherent Operator Based on $su(2)$}
Let us introduce generalized coherent operators and the states based on 
$su(2)$. Let $\{ j_{+}, j_{-}, j_{3} \}$ be a Weyl basis of Lie algebra 
$su(2) \subset sl(2,\fukuso)$, 
\begin{equation}
  \label{eq:2-3-1}
 j_{+} = \left(
        \begin{array}{cc}
               0 & 1 \\
               0 & 0 \\
        \end{array}
       \right),
 \quad 
 j_{-} = \left(
        \begin{array}{cc}
               0 & 0 \\
               1 & 0 \\
        \end{array}
       \right),
 \quad 
 j_{3} = \frac12 
       \left(
        \begin{array}{cc}
               1 &  0  \\
               0 & -1 \\
        \end{array}
       \right). 
\end{equation}
Then we have 
\begin{equation}
  \label{eq:2-3-2}
 [j_{3}, j_{+}]=j_{+}, \quad [j_{3}, j_{-}]=-j_{-}, 
 \quad [j_{+}, j_{-}]=2j_{3}.
\end{equation}
We note that $(j_{+})^{\dagger}=j_{-}$.

Next we consider a spin $J\ (> 0)$ representation of $su(2) 
\subset sl(2,\fukuso)$ and set its generators 
$\{ J_{+}, J_{-}, J_{3} \}\ ((J_{+})^{\dagger} = J_{-})$, 
\begin{equation}
  \label{eq:2-3-3}
 [J_{3}, J_{+}]=J_{+}, \quad [J_{3}, J_{-}]=-J_{-}, 
 \quad [J_{+}, J_{-}]=2J_{3}.
\end{equation}
We note that this (unitary) representation is necessarily finite dimensional. 
The Fock space on which $\{ J_{+}, J_{-}, J_{3} \}$ act is 
$\calh_{J} \equiv \{\ket{J,n} \vert 0 \le n \le 2J \}$ and 
whose actions are
\begin{eqnarray}
  \label{eq:2-3-4}
 J_{+} \ket{J,n} &=& \sqrt{(n+1)(2J-n)}\ket{J,n+1},\ 
 J_{-} \ket{J,n}  = \sqrt{n(2J-n+1)}\ket{J,n-1},  \nonumber  \\
 J_{3} \ket{J,n} &=& (-J+n)\ket{J,n}, 
\end{eqnarray}
where $\ket{J,0}$ is a normalized vacuum ($J_{-}\ket{J,0}=0$ and 
$\langle J,0|J,0 \rangle =1$). We have written $\ket{J,0}$ instead 
of $\ket{0}$ to emphasize the spin $J$ representation, see \cite{FKSF1}. 
From (\ref{eq:2-3-4}), states $\ket{J,n}$ are given by 
\begin{equation}
  \label{eq:2-3-5}
 \ket{J,n} =\frac{(J_{+})^n}{\sqrt{n!{}_{2J}P_n}}\ket{J,0}
\end{equation}
where ${}_{2J}P_n\equiv (2J)(2J-1)\cdots (2J-n+1)$.

\par \noindent
These states satisfy the orthonormality and completeness conditions 
\begin{equation}
  \label{eq:2-3-6}
  \langle J,m \vert J,n \rangle =\delta_{mn}, 
  \quad \sum_{n=0}^{2J}\ket{J,n}\bra{J,n}\ = \mathbf{1}_{J}.
\end{equation}
Now let us consider a generalized version of coherent states : 

\noindent{\bfseries Definition}\quad We call a state  
\begin{equation}
   \label{eq:2-3-7}
 \ket{z} = \mbox{e}^{zJ_{+} - \bar{z}J_{-}}\ket{J,0} 
      \equiv W(z)\ket{J,0}
  \quad \mbox{for} \quad z \in \fukuso.
\end{equation}
the generalized coherent state (or the coherent state of Perelomov's 
type based on $su(2)$ in our terminology).

\par \noindent  
We recall the disentangling formula :
\begin{equation}
  \label{eq:2-3-10}
    \mbox{e}^{zJ_{+} -\bar{z}J_{-}}  
  = \mbox{e}^{\eta J_{+}}\mbox{e}^{\log (1+\vert\eta\vert^2)J_{3}}
    \mbox{e}^{-\bar{\eta}J_{-}}
  = \mbox{e}^{-\bar{\eta}J_{-}}\mbox{e}^{-\log (1+\vert\eta\vert^2)J_{3}}
    \mbox{e}^{\eta J_{+}}, \quad 
    \eta \equiv \frac{\mbox{tan}(\zetta)}{\zetta}z.
\end{equation}
This is also the key formula for generalized coherent operators.

\par \vspace{3mm} \noindent
A comment is in order. 
We can construct the spin $K$ and $J$ representations by making 
use of Schwinger's boson method. But we don't repeat it here, see for 
example \cite{KF5}.

\section{Matrix Elements of Generalized Coherent Operators $\cdots$ 
Results}

In this section we first present matrix elements of coherent operators 
and next study matrix elements of generalized coherent operators. 

\par \noindent
Let us endeavor to make this section self--contained as far as we can 
because the proofs in the following are very important to understand 
mathematical structure of coherent or generalized coherent operators. 

\subsection{Matrix Elements of Coherent Operator}
We show explicit formulas of matrix elements of coherent operators in 
(\ref{eq:2-1-7}). This result is well--known, see for example \cite{AP}. 

\vspace{1cm}
\noindent{\bfseries  Matrix Elements}\quad The matrix elements of 
$U(z)$ are :
\begin{eqnarray}
   \label{eq:3-1-1-1}
 &&(\mbox{i})\quad n \le m \quad 
   \bra{n}U(z)\ket{m} = \mbox{e}^{-\frac{1}{2}\zetta^2}\sqrt{\frac{n!}{m!}}
                 (-\bar{z})^{m-n}{L_n}^{(m-n)}(\zetta^2), \\
   \label{eq:3-1-1-2}
 &&(\mbox{ii})\quad n \geq m \quad 
   \bra{n}U(z)\ket{m} = \mbox{e}^{-\frac{1}{2}\zetta^2}\sqrt{\frac{m!}{n!}}
                 z^{n-m}{L_m}^{(n-m)}(\zetta^2),
\end{eqnarray}
where ${L_n}^{(\alpha)}$ is the associated Laguerre polynomial defined by 
\begin{equation}
   \label{eq:3-1-2}
 {L_n}^{(\alpha)}(x)=\sum_{j=0}^{n}(-1)^j {{n+\alpha}\choose{n-j}}
                  \frac{x^j}{j!}. 
\end{equation}
In particular ${L_n}^{(0)} \equiv L_{n}$ is the usual Laguerre polynomial 
and these are related to diagonal elements of $U(z)$. 

\par \vspace{5mm} \noindent 
The proof is easy and as follows. 

For the case $n \geq m $ 
\begin{eqnarray}
 \bra{n}U(z)\ket{m} 
   &=& \frac{1}{\sqrt{n!}}\frac{1}{\sqrt{m!}}
     \bra{0}a^{n}\mbox{e}^{za^{\dagger}-{\bar z}a}(a^{\dagger})^{m}\ket{0} 
     \qquad  \quad \  \mbox{by (\ref{eq:2-1-3})}  \nonumber \\
   &=& \frac{1}{\sqrt{n! m!}}\mbox{e}^{\frac{1}{2}|z|^{2}}
     \bra{0}a^{n}\mbox{e}^{-{\bar z}a}\mbox{e}^{za^{\dagger}}
     (a^{\dagger})^{m}\ket{0} \qquad
             \mbox{by (\ref{eq:2-1-11})}  \nonumber \\
   &=& \frac{1}{\sqrt{n! m!}}\mbox{e}^{\frac{1}{2}|z|^{2}}
     \frac{\partial^{n}}{\partial(-{\bar z})^{n}}
     \frac{\partial^{m}}{\partial{z}^{m}}
     \bra{0}\mbox{e}^{-{\bar z}a}\mbox{e}^{za^{\dagger}}\ket{0}. \nonumber 
\end{eqnarray}
Noting 
\[
\mbox{e}^{-{\bar z}a}\mbox{e}^{za^{\dagger}}=
\mbox{e}^{-|z|^{2}}\mbox{e}^{za^{\dagger}}\mbox{e}^{-{\bar z}a}
\]
by (\ref{eq:2-1-11}) and $a\ket{0}=0$, $\bra{0}a^{\dagger}=0$
\begin{eqnarray}
 \bra{n}U(z)\ket{m} 
 &=&
     \frac{1}{\sqrt{n! m!}}\mbox{e}^{\frac{1}{2}|z|^{2}}
     \frac{\partial^{m}}{\partial{z}^{m}} 
     \frac{\partial^{n}}{\partial(-{\bar z})^{n}}
     \mbox{e}^{-|z|^{2}} \nonumber \\
 &=&
     \frac{1}{\sqrt{n! m!}}\mbox{e}^{\frac{1}{2}|z|^{2}}
     \frac{\partial^{m}}{\partial{z}^{m}} 
     \left(z^{n}\mbox{e}^{-|z|^{2}}\right) \nonumber  \\
 &=&
     \frac{1}{\sqrt{n! m!}}\mbox{e}^{\frac{1}{2}|z|^{2}}
     \sum_{k=0}^{m} {m\choose m-k} (z^{n})^{(m-k)}
     \left(\mbox{e}^{-|z|^{2}}\right)^{(k)}     \nonumber  \\
 &=&
     \frac{1}{\sqrt{n! m!}}\mbox{e}^{\frac{1}{2}|z|^{2}}
     \sum_{k=0}^{m} {m\choose m-k} n(n-1)\cdots (n-m+k+1)z^{n-m+k}
     (-{\bar z})^{k}\mbox{e}^{-|z|^{2}}    \nonumber  \\
 &=&
     \frac{1}{\sqrt{n! m!}}\mbox{e}^{-\frac{1}{2}|z|^{2}}
     z^{n-m}
     \sum_{k=0}^{m} (-1)^{k}\frac{m!}{k!(m-k)!} \frac{n!}{(n-m+k)!}
     (|z|^{2})^{k}  \nonumber  \\
 &=&
     \sqrt{\frac{m!}{n!}}\mbox{e}^{-\frac{1}{2}|z|^{2}}z^{n-m}
     \sum_{k=0}^{m} (-1)^{k}\frac{n!}{(m-k)!(n-m+k)!}
     \frac{(|z|^{2})^{k}}{k!}  \nonumber \\
 &=&
     \sqrt{\frac{m!}{n!}}\mbox{e}^{-\frac{1}{2}|z|^{2}}z^{n-m}
     \sum_{k=0}^{m} (-1)^{k} {m + n-m\choose m-k}
     \frac{(|z|^{2})^{k}}{k!}  \nonumber \\
 &=&
     \sqrt{\frac{m!}{n!}}\mbox{e}^{-\frac{1}{2}|z|^{2}}z^{n-m}
     {L_m}^{(n-m)}(|z|^{2}). \nonumber 
\end{eqnarray}
For the case $n \leq m $ we have only to take a complex conjugate of 
(\ref{eq:3-1-1-2}) (note that $U(z)^{\dagger}=U(-z)$). 

\par \vspace{5mm} 
Here we state an interesting application of matrix elements to the theory 
of special functions. From (\ref{eq:2-1-9})\quad
$
U(z+w) = \mbox{e}^{-\frac{1}{2}(z\bar{w}-\bar{z}w)}\ U(z)U(w) 
$
\quad 
let us take a matix element 
\begin{eqnarray}
\bra{n}U(z+w)\ket{m}&=&\mbox{e}^{-\frac{1}{2}(z\bar{w}-\bar{z}w)}
\bra{n}U(z)U(w)\ket{m}   \nonumber \\
&=&\mbox{e}^{-\frac{1}{2}(z\bar{w}-\bar{z}w)}
\sum_{k=0}^{\infty} \bra{n}U(z)\ket{k} \bra{k}U(w)\ket{m}. \nonumber 
\end{eqnarray}
Then by substituting (\ref{eq:3-1-1-1}) and (\ref{eq:3-1-1-2}) 
into the above equation and making some calculations 
we obtain 
\begin{flushleft}
(i) $N \geq 1$ 
\end{flushleft}
\begin{eqnarray}
\label{eq:N>1}
L_{m}^{(N)}(|z+w|^2)
=\mbox{e}^{\bar{z}w}&&
\left\{
\left(\frac{z}{z+w}\right)^{N}\sum_{k=0}^{m} \frac{k!}{m!}(-z\bar{w})^{m-k}
L_{k}^{(m+N-k)}(|z|^2)L_{k}^{(m-k)}(|w|^2)
\right.  \nonumber \\
&&+
\left(\frac{1}{z+w}\right)^{N}\sum_{k=m+1}^{m+N}z^{m+N-k}w^{k-m}
L_{k}^{(m+N-k)}(|z|^2)L_{m}^{(k-m)}(|w|^2) \nonumber \\
&&+
\left.
\left(\frac{w}{z+w}\right)^{N}\sum_{k=m+N+1}^{\infty} \frac{(m+N)!}{k!}
(-\bar{z}w)^{k-m-N}L_{m+N}^{(k-m-N)}(|z|^2)L_{m}^{(k-m)}(|w|^2)
\right\},  \nonumber \\
&&{}
\end{eqnarray}
\begin{flushleft}
(ii) $N = 0$ 
\end{flushleft}
\begin{eqnarray}
\label{eq:N=0}
L_{m}(|z+w|^2)
=\mbox{e}^{\bar{z}w}&&
\left\{\quad 
\sum_{k=0}^{m} \frac{k!}{m!}(-z\bar{w})^{m-k}
L_{k}^{(m-k)}(|z|^2)L_{k}^{(m-k)}(|w|^2)
\right.  \nonumber \\
&&+
\left.
\sum_{k=m+1}^{\infty} \frac{m!}{k!}
(-\bar{z}w)^{k-m}L_{m}^{(k-m)}(|z|^2)L_{m}^{(k-m)}(|w|^2)
\right\}. 
\end{eqnarray}
To obtain equations of these types is not our purpose of this paper, so 
we omit the proof (we leave it to the readers). 

Since the equation (\ref{eq:2-1-9}) is based on the elementary 
Baker-Campbell-Hausdorff formula (\ref{eq:2-1-12}), our equations 
(\ref{eq:N>1}) and (\ref{eq:N=0}) are highly non--trivial. 
What is the mathematical meaning of these formulas ? 
We think they are a kind of factorization formula.

\subsection{Matrix Elements of Coherent Operator Based on $su(1,1)$}

We in this section study matrix elements of the generalized coherent 
operator based on Lie algebra $su(1,1)$ (\ref{eq:2-2-8}). 

\vspace{1cm}
\noindent{\bfseries  Matrix Elements}\quad The matrix elements of 
$V(z)$ are :
\begin{eqnarray}
   \label{eq:3-2-1-1}
  (\mbox{i})\quad n \le m \quad 
   && \bra{K,n}V(z)\ket{K,m}= 
    \sqrt{\frac{n!m!}{(2K)_n(2K)_m}}
    \frac{(-\bar{\kappa})^{m-n}}{(1+\kappazetta^2)^{K+\frac{n+m}{2}}}
    \ \times   \nonumber \\
   &&\sum_{j=0}^{n}(-1)^{n-j}\frac{\Gamma(2K+m+n-j)}{\Gamma(2K)(m-j)!
     (n-j)!j!}(1+\kappazetta^2)^j(\kappazetta^2)^{n-j}, \\ 
   \label{eq:3-2-1-2}
  (\mbox{ii})\quad n \geq m \quad    
    && \bra{K,n}V(z)\ket{K,m}=
    \sqrt{\frac{n!m!}{(2K)_n(2K)_m}}
    \frac{\kappa^{n-m}}{(1+\kappazetta^2)^{K+\frac{n+m}{2}}}
    \ \times   \nonumber \\
   &&\sum_{j=0}^{m}(-1)^{m-j}\frac{\Gamma(2K+m+n-j)}{\Gamma(2K)(m-j)!
     (n-j)!j!}(1+\kappazetta^2)^j(\kappazetta^2)^{m-j}, 
\end{eqnarray}
where 
\begin{equation}
   \label{eq:3-2-2}
\kappa \equiv \frac{\mbox{sinh}(\zetta)}{\zetta}z 
       ={\mbox{cosh}(\zetta)}\zeta.
\end{equation}

{\bf A comment is in order}. The author doesn't know whether or not 
the right  hand sides of (\ref{eq:3-2-1-1}) and (\ref{eq:3-2-1-2}) 
could be expressed by some known special functions such as 
associated Laguerre functions in (\ref{eq:3-1-2}).

\par \vspace{5mm} \noindent 
This result has been reported in \cite{KF5} under some assumption
($2K \in \futon$). 
In this paper we remove this extra condition and give a complete 
proof to this result. 

\par \noindent 
Since (\ref{eq:3-2-1-1}) is given by taking a complex conjugate of 
(\ref{eq:3-2-1-2}), we have only to prove (\ref{eq:3-2-1-2}).  
The proof is as follows.
\begin{flushleft}
{\bf Step 1}\quad For the case $n \geq m $ 
\end{flushleft}
\begin{eqnarray}
&&\bra{K,n}V(z)\ket{K,m} \nonumber \\
   &=& \frac{1}{\sqrt{n!(2K)_{n}}}\frac{1}{\sqrt{m!(2K)_{m}}}
     \bra{K,0}{K_{-}}^{n}\mbox{e}^{zK_{+}-{\bar z}K_{-}}
     {K_{+}}^{m}\ket{K,0} \qquad \qquad \qquad \ 
             \mbox{by (\ref{eq:2-2-5})}  \nonumber \\
&=& \frac{1}{\sqrt{n!m!(2K)_{n}(2K)_{m}}} 
     \bra{K,0}{K_{-}}^{n}
      \mbox{e}^{-{\bar \zeta}K_{-}}
      \mbox{e}^{-\log(1-|\zeta|^{2})K_{3}}
      \mbox{e}^{\zeta K_{+}}
     {K_{+}}^{m}\ket{K,0}\ \quad 
             \mbox{by (\ref{eq:2-2-11})}  \nonumber \\
&=& \frac{1}{\sqrt{n!m!(2K)_{n}(2K)_{m}}} 
    \frac{\partial^{n}}{\partial \alpha^{n}}
    \frac{\partial^{m}}{\partial \beta^{m}}
     \bra{K,0}{
      \mbox{e}^{(\alpha-{\bar \zeta})K_{-}}
      \mbox{e}^{-\log(1-|\zeta|^{2})K_{3}}
      \mbox{e}^{(\beta+\zeta)K_{+}}
              }\ket{K,0}|_{\alpha=\beta=0} \nonumber \\
\label{eq:first-stage}
&=& \frac{1}{\sqrt{n!m!(2K)_{n}(2K)_{m}}} 
    \frac{\partial^{n}}{\partial \alpha^{n}}
    \frac{\partial^{m}}{\partial \beta^{m}}
    \left\{
       (1-|\zeta|^2)^{1/2}-
       \frac{(\alpha-{\bar \zeta})(\beta+\zeta)}{(1-|\zeta|^2)^{1/2}}
    \right\}^{-2K}|_{\alpha=\beta=0},
\end{eqnarray}
where we have used the exchange relation like (\ref{eq:2-1-11}) :  
\begin{flushleft}
\begin{Large}
{\bf Formula}
\end{Large}
\end{flushleft}
\begin{equation}
\mbox{e}^{aK_{-}}\mbox{e}^{2bK_{3}}\mbox{e}^{cK_{+}}=
\mbox{e}^{xK_{+}}\mbox{e}^{2yK_{3}}\mbox{e}^{zK_{-}}
\end{equation}
where 
\begin{equation}
x=\frac{c\mbox{e}^{b}}{\mbox{e}^{-b}-ac\mbox{e}^{b}},\quad 
y=-{\log}(\mbox{e}^{-b}-ac\mbox{e}^{b}),\quad 
z=\frac{a\mbox{e}^{b}}{\mbox{e}^{-b}-ac\mbox{e}^{b}}.
\end{equation}
For the proof see Appendix. 

\vspace{5mm}
\begin{flushleft}
{\bf Step 2}\quad Let us calculate the differential \quad 
\end{flushleft}
\begin{eqnarray}
\label{eq:step-2}
  &&\frac{\partial^{n}}{\partial \alpha^{n}}
    \frac{\partial^{m}}{\partial \beta^{m}}
    \left\{
       (1-|\zeta|^2)^{1/2}-
       \frac{(\alpha-{\bar \zeta})(\beta+\zeta)}{(1-|\zeta|^2)^{1/2}}
    \right\}^{-2K}|_{\alpha=\beta=0}  \nonumber \\
=&&
      \left\{
            \frac{1}{(1-|\zeta|^2)^{1/2}}
      \right\}^{-2K}
    \frac{\partial^{m}}{\partial \beta^{m}}
    \frac{\partial^{n}}{\partial \alpha^{n}}
    \left\{
       1-|\zeta|^2-(\alpha-{\bar \zeta})(\beta+\zeta)
    \right\}^{-2K}|_{\alpha=\beta=0}  \nonumber \\
  =&&
      \left\{
            \frac{1}{(1-|\zeta|^2)^{1/2}}
      \right\}^{-2K}  
  \frac{\partial^{m}}{\partial \beta^{m}}
    (2K)_{n}
    (\beta+\zeta)^{n}
    \left\{
       1-|\zeta|^2+{\bar \zeta}(\beta+\zeta)
    \right\}^{-2K-n}|_{\beta=0}  \nonumber \\
  =&&(2K)_{n}
      \left\{
            \frac{1}{(1-|\zeta|^2)^{1/2}}
      \right\}^{-2K}
    \frac{\partial^{m}}{\partial \beta^{m}}
    \left\{
    (\beta+\zeta)^{n}(1+\beta{\bar \zeta})^{-2K-n}
    \right\}|_{\beta=0}  \nonumber \\
  =&&(2K)_{n}
      \left\{
            \frac{1}{(1-|\zeta|^2)^{1/2}}
      \right\}^{-2K}
  \sum_{j=0}^{m}{m\choose j}
    \left\{(\beta+\zeta)^{n}\right\}^{(j)}
    \left\{(1+\beta{\bar \zeta})^{-2K-n}\right\}^{(m-j)}
    |_{\beta=0}  \nonumber \\
  =&&(2K)_{n}
      \left\{
            \frac{1}{(1-|\zeta|^2)^{1/2}}
      \right\}^{-2K}
   \sum_{j=0}^{m} {m\choose j}\frac{n!}{(n-j)!}{\zeta}^{n-j}
           (-1)^{m-j}(2K+n)_{m-j}{\bar \zeta}^{m-j} 
                  \nonumber \\
  =&&
      \left\{
            \frac{1}{(1-|\zeta|^2)^{1/2}}
      \right\}^{-2K}{\zeta}^{n-m}
   \sum_{j=0}^{m}(-1)^{m-j} {m\choose j}\frac{n!}{(n-j)!}
           (2K)_{n}(2K+n)_{m-j}(|\zeta|^2)^{m-j}
                  \nonumber \\
\label{eq:middle-stage}
  =&&
      \left\{
            \frac{1}{(1-|\zeta|^2)^{1/2}}
      \right\}^{-2K-n-m}
      \left(\frac{\zeta}{(1-|\zeta|^2)^{1/2}}\right)^{n-m}\times 
        \nonumber \\
 &&\qquad \quad  
       \sum_{j=0}^{m}(-1)^{m-j} {m\choose j}\frac{n!}{(n-j)!}
       (2K)_{n+m-j}
       \left(\frac{|\zeta|^2}{1-|\zeta|^2}\right)^{m-j}
       \left(\frac{1}{1-|\zeta|^2}\right)^{j}.
\end{eqnarray}
Now let us make a change of variables 
\[
  \zeta \longrightarrow \kappa \equiv \frac{\zeta}{\sqrt{1-|\zeta|^{2}}} 
  \quad 
\]
then by (\ref{eq:2-2-11})
\[
 \kappa=\frac{z\sinh(|z|)}{|z|},\quad |\kappa|^{2}=\sinh^{2}(|z|),
 \quad 1+|\kappa|^{2}=\cosh^{2}(|z|)=\frac{1}{1-|\zeta|^{2}}.
\]
Therefore 
\begin{eqnarray}
\mbox{(\ref{eq:middle-stage})}
&=&
      (1+|\kappa|^2)^{-(K+\frac{n+m}{2})}
      \kappa^{n-m}
      \sum_{j=0}^{m}(-1)^{m-j} {m\choose j}\frac{n!}{(n-j)!}
      (2K)_{n+m-j}(|\kappa|^2)^{m-j}(1+|\kappa|^2)^{j} \nonumber \\
&=&n!m!\kappa^{n-m}(1+|\kappa|^2)^{-(K+\frac{n+m}{2})}
      \sum_{j=0}^{m}(-1)^{m-j} \frac{(2K)_{n+m-j}}{j!(m-j)!(n-j)!}
      (|\kappa|^2)^{m-j}(1+|\kappa|^2)^{j} \nonumber \\
\label{eq:last-stage}
&=&n!m!\kappa^{n-m}(1+|\kappa|^2)^{-(K+\frac{n+m}{2})}
   \sum_{j=0}^{m}(-1)^{m-j} \frac{\Gamma(2K+n+m-j)}{\Gamma(2K)j!(m-j)!(n-j)!}
   (|\kappa|^2)^{m-j}(1+|\kappa|^2)^{j}, \nonumber \\
&&{}
\end{eqnarray}
where we have used the formula
\[
  (2K)_{j}\equiv (2K)(2K+1)\cdots (2K+j-1)=\frac{\Gamma(2K+j)}{\Gamma(2K)}.
\]
Combining (\ref{eq:first-stage}) with (\ref{eq:last-stage}) we finally obtain 
the matrix elements (\ref{eq:3-2-1-2}).

\subsection{Matrix Elements of Coherent Operator Based on $su(2)$}

We in this section study matrix elements of the generalized coherent 
operator based on Lie algebra $su(2)$\ (\ref{eq:2-3-7}). 
In this case it is always $2J\in \futon$. 

\noindent{\bfseries  Matrix Elements}\quad The matrix elements of 
$W(z)$ are :
\begin{eqnarray}
   \label{eq:3-3-1-1}
  (\mbox{i})\quad n \le m \quad 
   && \bra{J,n}W(z)\ket{J,m}= 
    \sqrt{\frac{n!m!}{{}_{2J}P_n {}_{2J}P_m}}
    (-\bar{\kappa})^{m-n}(1-\kappazetta^2)^{J-\frac{n+m}{2}}
    \ \times   \nonumber \\
   &&\sum_{j=0}^{n}{}_{*}(-1)^{n-j}\frac{(2J)!}{(2J-m-n+j)!(m-j)!(n-j)!j!}
     (1-\kappazetta^2)^{j}(\kappazetta^2)^{n-j},\quad    \\ 
   \label{eq:3-3-1-2}
  (\mbox{ii})\quad n \geq m \quad    
    && \bra{J,n}W(z)\ket{J,m}=
    \sqrt{\frac{n!m!}{{}_{2J}P_n {}_{2J}P_m}}
    \kappa^{n-m}(1-\kappazetta^2)^{J-\frac{n+m}{2}}
    \ \times   \nonumber \\
   &&\sum_{j=0}^{m}{}_{*}(-1)^{m-j}\frac{(2J)!}{(2J-m-n+j)!(m-j)!(n-j)!j!}
     (1-\kappazetta^2)^{j}(\kappazetta^2)^{m-j},\quad  
\end{eqnarray}
where 
\begin{equation}
   \label{eq:3-3-2}
\kappa \equiv \frac{\mbox{sin}(\zetta)}{\zetta}z 
       ={\mbox{cos}(\zetta)}\eta.
\end{equation}
Here $\sum{}_{*}$ means a summation over $j$ satisfying $2J-m-n+j \geq 0$. 

\par \noindent
{\bf A comment is in order}. The author doesn't know whether or not the right  
hand sides of (\ref{eq:3-3-1-1}) and (\ref{eq:3-3-1-2}) could be expressed 
in terms of some known special functions.

The proof is almost the same as the preceding one, so we leave the proof of 
the first step to the readers. 
For the case $n \geq m$ we reach 
\begin{eqnarray}
\label{eq:J-step-2}
\bra{J,n}W(z)\ket{J,m}
  =&&\frac{1}{\sqrt{n!m! {}_{2J}P_n {}_{2J}P_m}}
    \frac{\partial^{n}}{\partial \alpha^{n}}
    \frac{\partial^{m}}{\partial \beta^{m}}
    \left\{
       (1+|\eta|^2)^{1/2}+
       \frac{(\alpha-{\bar \eta})(\beta+\eta)}{(1+|\eta|^2)^{1/2}}
    \right\}^{2J}|_{\alpha=\beta=0}.  \nonumber \\
&&{}
\end{eqnarray}

\par \vspace{5mm} \noindent
Let us calculate the differential.
\begin{eqnarray}
\label{eq:step-2-J}
  &&\frac{\partial^{n}}{\partial \alpha^{n}}
    \frac{\partial^{m}}{\partial \beta^{m}}
    \left\{
       (1+|\eta|^2)^{1/2}+
       \frac{(\alpha-{\bar \eta})(\beta+\eta)}{(1+|\eta|^2)^{1/2}}
    \right\}^{2J}|_{\alpha=\beta=0}  \nonumber \\
  &=&
    \left\{
    \frac{1}{(1+|\eta|^2)^{1/2}}
    \right\}^{2J}
    \frac{\partial^{m}}{\partial \beta^{m}}
    \frac{\partial^{n}}{\partial \alpha^{n}}
   \left\{
       1+|\eta|^2+(\alpha-{\bar \eta})(\beta+\eta)
    \right\}^{2J}|_{\alpha=\beta=0}  \nonumber \\
  &=&
    \left\{
    \frac{1}{(1+|\eta|^2)^{1/2}}
    \right\}^{2J}
  \frac{\partial^{m}}{\partial \beta^{m}}
    {}_{2J}P_n (\beta+\eta)^{n}
    \left\{
       1+|\eta|^2-{\bar \eta}(\beta+\eta)
    \right\}^{2J-n}|_{\beta=0}  \nonumber \\
  &=&{}_{2J}P_n
      \left\{
            \frac{1}{(1+|\eta|^2)^{1/2}}
      \right\}^{2J}
    \frac{\partial^{m}}{\partial \beta^{m}}
    \left\{
    (\beta+\eta)^{n}(1-\beta{\bar \eta})^{2J-n}
    \right\}|_{\beta=0}  \nonumber \\
  &=&{}_{2J}P_n
      \left\{
            \frac{1}{(1+|\eta|^2)^{1/2}}
      \right\}^{2J}
  \sum_{j=0}^{m}{}_{*}{m\choose j}
    \left\{(\beta+\eta)^{n}\right\}^{(j)}
    \left\{(1-\beta{\bar \eta})^{2J-n}\right\}^{(m-j)}
    |_{\beta=0}  \nonumber \\
  &=&{}_{2J}P_n
      \left\{
            \frac{1}{(1+|\eta|^2)^{1/2}}
      \right\}^{2J}
   \sum_{j=0}^{m}{}_{*} {m\choose j}\frac{n!}{(n-j)!}{\eta}^{n-j}
           {}_{2J-n}P_{m-j}(-{\bar \eta})^{m-j} 
                  \nonumber \\
  &=&
      \left\{
            \frac{1}{(1+|\eta|^2)^{1/2}}
      \right\}^{2J}{\eta}^{n-m}
   \sum_{j=0}^{m}{}_{*} (-1)^{m-j} {m\choose j}\frac{n!}{(n-j)!}
          {}_{2J}P_n\ {}_{2J-n}P_{m-j}(|\eta|^2)^{m-j}
                  \nonumber \\
\label{eq:middle-stage-J}
  &=&
      \left\{
            \frac{1}{(1+|\eta|^2)^{1/2}}
      \right\}^{2J-n-m}
      \left(\frac{\eta}{(1+|\eta|^2)^{1/2}}\right)^{n-m}\times 
        \nonumber \\
 &&\qquad \quad  
       \sum_{j=0}^{m}{}_{*} (-1)^{m-j} {m\choose j}\frac{n!}{(n-j)!}
       {}_{2J}P_{n+m-j}
       \left(\frac{|\eta|^2}{1+|\eta|^2}\right)^{m-j}
       \left(\frac{1}{1+|\eta|^2}\right)^{j} \nonumber \\
&=&
      (1-|\kappa|^2)^{J-\frac{n+m}{2}}
      \kappa^{n-m}
      \sum_{j=0}^{m}{}_{*} (-1)^{m-j} {m\choose j}\frac{n!}{(n-j)!}
      {}_{2J}P_{n+m-j}(|\kappa|^2)^{m-j}(1-|\kappa|^2)^{j} \nonumber \\
&=&n!m!\kappa^{n-m}(1-|\kappa|^2)^{J-\frac{n+m}{2}}
  \sum_{j=0}^{m}{}_{*} (-1)^{m-j} 
  \frac{{}_{2J}P_{n+m-j}}{j!(m-j)!(n-j)!}
  (|\kappa|^2)^{m-j}(1-|\kappa|^2)^{j} \nonumber \\
\label{eq:last-stage-J}
&=&n!m!\kappa^{n-m}(1-|\kappa|^2)^{J-\frac{n+m}{2}}\times 
  \nonumber \\
 &&\qquad \quad  
  \sum_{j=0}^{m}{}_{*} (-1)^{m-j} \frac{(2J)!}{(2J-n-m+j)!j!(m-j)!(n-j)!}
   (|\kappa|^2)^{m-j}(1-|\kappa|^2)^{j},
\end{eqnarray}
where we have used a change of variables 
\[
  \eta \longrightarrow \kappa \equiv \frac{\eta}{\sqrt{1+|\eta|^{2}}} 
  \quad \Longrightarrow \quad 
  |\kappa|^{2}=\frac{|\eta|^2}{1+|\eta|^2},\quad 
  1-|\kappa|^{2}=\frac{1}{1+|\eta|^2}
\]
and the formula 
\[
  {}_{2J}P_{j}=(2J)(2J-1)\cdots (2J-j+1)=\frac{(2J)!}{(2J-j)!}\ .
\]
Combining (\ref{eq:J-step-2}) with (\ref{eq:last-stage-J}) we finally obtain 
the matrix elements (\ref{eq:3-3-1-2}).

\section{A Problem}

In this section let us present a problem for the readers. 

In the definition ((\ref{eq:2-3-7}), (\ref{eq:2-2-8}))
of generalized coherent operators based on Lie 
algebras $su(2)$ and $su(1,1)$ \ operators $J_3$ and $K_3$ are not used. 
To use these ones we would like to extend generalized coherent operators 
as follows (see \cite{KF4}). 

\noindent{\bfseries Definition}\quad We set 
\begin{eqnarray}
  \label{eq:4-1}
  su(2) &:&\quad 
W(z,t) = \mbox{e}^{zJ_{+} - \bar{z}J_{-} + 2itJ_{3}}\quad 
  {\rm for}\  z \in \fukuso ,\ t \in \real \\
  \label{eq:4-2}
  su(1,1) &:&\quad 
V(z,t) = \mbox{e}^{zK_{+} - \bar{z}K_{-} + 2itK_{3}}\quad 
  {\rm for}\  z \in \fukuso, \ t \in \real.
\end{eqnarray}
Both $W(z,t)$ and $V(z,t)$ are unitary and $W(z,0)=W(z)$, $V(z,0)=V(z)$. 
On the other hand 
we have used operators of these types in the process of proof, so it is 
very natural to consider the above. Then 
\begin{flushleft}
{\bf Problem}\quad Determine the matrix elements of $W(z,t)$ and $V(z,t)$ :
\end{flushleft}
\begin{eqnarray}
  \label{eq:4-3-1}
  su(2) &:&\quad 
  \bra{J,n}W(z,t)\ket{J,m} \quad 
  {\rm for}\quad 0 \leq n,\ m \leq 2J , \\
  \label{eq:4-3-2}
  su(1,1) &:&\quad 
  \bra{K,n}V(z,t)\ket{K,m} \quad 
  {\rm for}\quad 0 \leq n,\ m . 
\end{eqnarray}
See for example \cite{KF6}.

\section{Discussion}

In this paper we have determined the matrix elements of the generalized 
coherent operators based on $su(1,1)$ and $su(2)$ in a perfect manner. 

By the way we have an interesting application of these matrix elements. 
In \cite{MFr} Frasca has used matrix elements of coherent operator in 
section 3.1 to explain the recent experimental results on 
Rabi oscillations in a Josephson junction \cite{NPT}. 
See also \cite{MFr2}. 

Since we have calculated matrix elements of generalized coherent operators 
we should generalize his method. 
We will report the results in a forthcoming paper \cite{KF15}.

We believe strongly that our calculations will play an essential role 
in understanding the general (mathematical) structure of Rabi oscillations 
in the strong coupling regime.

\par \vspace{10mm}

\noindent{\em Acknowledgment.}\\
The author wishes to thank K. Funahashi and R. Sasaki for their helpful 
comments and suggestions, and also Y. Machida for his warm hospitality at 
Numazu College of Technology.

\par \vspace{10mm}
\begin{center}
 \begin{Large}
   {\bf Appendix}
 \end{Large}
\end{center}
\vspace{5mm}
\begin{flushleft}
\begin{Large}
{\bf An Exchange Relation of Operators}
\end{Large}
\end{flushleft}
Let us prove the exchange relation (\ref{eq:fundamental-formula-su(1,1)}). 
\begin{flushleft}
{\bf Formula}
\end{flushleft}
\begin{equation}
\label{eq:fundamental-formula-su(1,1)}
\mbox{e}^{aK_{-}}\mbox{e}^{2bK_{3}}\mbox{e}^{cK_{+}}=
\mbox{e}^{xK_{+}}\mbox{e}^{2yK_{3}}\mbox{e}^{zK_{-}}
\end{equation}
where 
\begin{equation}
x=\frac{c\mbox{e}^{b}}{\mbox{e}^{-b}-ac\mbox{e}^{b}},\quad 
y=-{\log}(\mbox{e}^{-b}-ac\mbox{e}^{b}),\quad 
z=\frac{a\mbox{e}^{b}}{\mbox{e}^{-b}-ac\mbox{e}^{b}}.
\end{equation}

\par \vspace{5mm} \noindent
We divide the proof into two parts. 

First we assume that $\{ K_{+}, K_{-}, K_{3} \}$ is a differential 
representation of Lie group $SU(1,1)$, namely we have a unitary 
representation  
\[
\rho : SU(1,1) \subset SL(2, \fukuso) \longrightarrow U(\calh\otimes \calh) 
\]
such that 
\[
K_{-}=d\rho(k_{-}),\quad K_{3}=d\rho(k_{3}),\quad K_{+}=d\rho(k_{+}).
\]
In this case we must take $2K\in \futon$ (an integrability condition), 
see \cite{KF5}. Then since $\rho$ is a homomorphism of the {\bf group} 
\begin{eqnarray}
\mbox{e}^{aK_{-}}\mbox{e}^{2bK_{3}}\mbox{e}^{cK_{+}}
&=&\mbox{e}^{a d\rho(k_{-})}
 \mbox{e}^{2b d\rho(k_{3})}
 \mbox{e}^{c d\rho(k_{+})}
=\rho(\mbox{e}^{a k_{-}})
 \rho(\mbox{e}^{2b k_{3}})
 \rho(\mbox{e}^{c k_{+}})     \nonumber \\
&=&\rho(\mbox{e}^{a k_{-}}\mbox{e}^{2b k_{3}}\mbox{e}^{c k_{+}})
\nonumber \\
&=&\rho\left(
\mbox{exp}
\left(
  \begin{array}{cc}
       &  0 \\
     -a&  
  \end{array}
\right)
\mbox{exp}
\left(
  \begin{array}{cc}
      b&   \\
       & -b
  \end{array}
\right)
\mbox{exp}
\left(
  \begin{array}{cc}
       &  c \\
      0&  
  \end{array}
\right)
\right)       \nonumber \\
&=&\rho\left(
\left(
  \begin{array}{cc}
      1&  0 \\
     -a&  1
  \end{array}
\right)
\left(
  \begin{array}{cc}
    \mbox{e}^{b}&  \\
     &  \mbox{e}^{-b}
  \end{array}
\right)
\left(
  \begin{array}{cc}
      1&  c \\
      0& 1
  \end{array}
\right)
\right)   \nonumber \\
&=&\rho\left(
\left(
  \begin{array}{cc}
        \mbox{e}^{b}&  c\mbox{e}^{b}                \\
      -a\mbox{e}^{b}&  \mbox{e}^{-b}-ac\mbox{e}^{b}
  \end{array}
\right)
\right)\equiv \rho(A).  \nonumber 
\end{eqnarray}
Next let us make another Gauss decomposition of $A$. It is easy to see 
\[
A=
\left(
  \begin{array}{cc}
      1&  \frac{c\mbox{e}^{b}}{\mbox{e}^{-b}-ac\mbox{e}^{b}} \\
      0&  1
  \end{array}
\right)
\left(
  \begin{array}{cc}
      \frac{1}{\mbox{e}^{-b}-ac\mbox{e}^{b}}&   \\
       & \mbox{e}^{-b}-ac\mbox{e}^{b}
  \end{array}
\right)
\left(
  \begin{array}{cc}
      1&   0 \\
      -\frac{a\mbox{e}^{b}}{\mbox{e}^{-b}-ac\mbox{e}^{b}}&  1
  \end{array}
\right). 
\]
Here we set $f= \mbox{e}^{-b}-ac\mbox{e}^{b}$ for simplicity. Therefore 
\begin{eqnarray}
\rho(A)  
&=&\rho\left(
\left(
  \begin{array}{cc}
      1&  \frac{c\mbox{e}^{b}}{f} \\
      0&  1
  \end{array}
\right)
    \right)
\rho\left(
\left(
  \begin{array}{cc}
      \frac{1}{f}&   \\
                 & f
  \end{array}
\right)
    \right)
\rho\left(
\left(
  \begin{array}{cc}
      1&   0 \\
      -\frac{a\mbox{e}^{b}}{f}&  1
  \end{array}
\right)
     \right)  \nonumber \\
&=&\rho\left(
\mbox{exp}
\left(
  \begin{array}{cc}
      0&  \frac{c\mbox{e}^{b}}{f} \\
      0&  0
  \end{array}
\right)
       \right)
\rho\left(
\mbox{exp}
\left(
  \begin{array}{cc}
     -{\log}(f)&    \\
      &  {\log}(f)
  \end{array}
\right)
       \right)
\rho\left(
\mbox{exp}
\left(
  \begin{array}{cc}
      0&  0 \\
      -\frac{a\mbox{e}^{b}}{f}&  0
  \end{array}
\right) 
      \right) \nonumber \\
&=&
\rho\left(\mbox{e}^{\frac{c\mbox{e}^{b}}{f} k_{+}}\right)
\rho\left(\mbox{e}^{-2{\log}(f) k_{3}}\right)
\rho\left(\mbox{e}^{\frac{a\mbox{e}^{b}}{f} k_{-}}\right)
\nonumber \\
&=&
\mbox{e}^{\frac{c\mbox{e}^{b}}{f} d\rho(k_{+})}
\mbox{e}^{-2{\log}(f) d\rho(k_{3})}
\mbox{e}^{\frac{a\mbox{e}^{b}}{f} d\rho(k_{-})}
\nonumber \\
&=&
\mbox{e}^{\frac{c\mbox{e}^{b}}{f} K_{+}}
\mbox{e}^{-2{\log}(f) K_{3}}
\mbox{e}^{\frac{a\mbox{e}^{b}}{f} K_{-}}. 
\end{eqnarray}
That is, we obtained the formula under the condition $2K\in \futon$.  
Next to remove this condition we use a method of differential equations. 
We set  
\begin{eqnarray}
\label{eq:function-f}
F(a)&=&\mbox{e}^{aK_{-}}\mbox{e}^{2bK_{3}}\mbox{e}^{cK_{+}},   \\
\label{eq:function-g}
G(a)&=&
\mbox{e}^{\frac{c \mbox{e}^{b}}{f(a)} K_{+}}
\mbox{e}^{-2{\log}f(a) K_{3}}
\mbox{e}^{\frac{a \mbox{e}^{b}}{f(a)} K_{-}},
\end{eqnarray}
where $f(a)=\mbox{e}^{-b}-ac\mbox{e}^{b}$. 

\par \noindent
First 
\begin{equation}
F(0)=\mbox{e}^{2b K_{3}}\mbox{e}^{c K_{+}},\quad 
F{'}(a)=K_{-}F(a). 
\end{equation}
Next by (\ref{eq:function-g}) 
\begin{equation}
G(0)=\mbox{e}^{c \mbox{e}^{2b} K_{+}}\mbox{e}^{2b K_{3}}
\end{equation}
so 
\begin{eqnarray}
G(0)&=&\mbox{e}^{c \mbox{e}^{2b} K_{+}}\mbox{e}^{2b K_{3}}
=
\mbox{e}^{2b K_{3}}\mbox{e}^{-2b K_{3}}
\mbox{e}^{c \mbox{e}^{2b} K_{+}}\mbox{e}^{2b K_{3}}
=
\mbox{e}^{2b K_{3}}
\mbox{e}^{c \mbox{e}^{2b} \mbox{e}^{-2b K_{3}}K_{+}\mbox{e}^{2b K_{3}} }    
\nonumber \\
&=&
\mbox{e}^{2b K_{3}}
\mbox{e}^{c \mbox{e}^{2b} \mbox{e}^{-2b}K_{+} }   
=
\mbox{e}^{2b K_{3}}\mbox{e}^{c K_{+} } = F(0).
\end{eqnarray}
Here we have used the formula 
\[
\mbox{e}^{\alpha K_{3}}K_{+}\mbox{e}^{-\alpha K_{3}}=
\mbox{e}^{\alpha}K_{+}. 
\]
Moreover 
\begin{eqnarray}
\label{eq:continuation}
&&G{'}(a)  \nonumber \\
=&&
\frac{(c\mbox{e}^{b})^{2}}{f(a)^{2}}
K_{+}G(a)
+
\frac{2c\mbox{e}^{b}}{f(a)}
\mbox{e}^{\frac{c \mbox{e}^{b}}{f(a)} K_{+}}
K_{3}\mbox{e}^{-2{\log}f(a) K_{3}}
\mbox{e}^{\frac{a \mbox{e}^{b}}{f(a)} K_{-}}
+
\frac{1}{f(a)^{2}}G(a)K_{-}        \nonumber \\
=&&
\frac{(c\mbox{e}^{b})^{2}}{f(a)^{2}}
K_{+}G(a)
+
\frac{2c\mbox{e}^{b}}{f(a)}
\left\{
\mbox{e}^{\frac{c \mbox{e}^{b}}{f(a)} K_{+}}
K_{3}
\mbox{e}^{-\frac{c \mbox{e}^{b}}{f(a)} K_{+}}
\right\}G(a)
+
\frac{1}{f(a)^{2}}\left\{G(a)K_{-}G(a)^{-1}\right\}G(a)  \nonumber \\
=&&
\frac{(c\mbox{e}^{b})^{2}}{f(a)^{2}}
K_{+}G(a)
+
\frac{2c\mbox{e}^{b}}{f(a)}
\left(
K_{3}-\frac{c \mbox{e}^{b}}{f(a)}K_{+}
\right)G(a)
+
\frac{1}{f(a)^{2}}\left\{G(a)K_{-}G(a)^{-1}\right\}G(a) \nonumber \\
=&&\frac{1}{f(a)^{2}}
\left\{
(c\mbox{e}^{b})^{2}K_{+}
+
2c\mbox{e}^{b}f(a)
\left(
K_{3}-\frac{c \mbox{e}^{b}}{f(a)}K_{+}
\right)
+
G(a)K_{-}G(a)^{-1}
\right\}G(a).
\end{eqnarray}
Here we have used the formula 
\[
\mbox{e}^{\alpha K_{+}}K_{3}\mbox{e}^{-\alpha K_{+}}=
K_{3}-\alpha K_{+}. 
\]
Now let us calculate the term $G(a)K_{-}G(a)^{-1}$ : 
\begin{eqnarray}
\label{eq:continuation-2}
G(a)K_{-}G(a)^{-1}
&=&
\mbox{e}^{\frac{c \mbox{e}^{b}}{f(a)} K_{+}}
\left\{
\mbox{e}^{-2{\log}f(a) K_{3}}
K_{-}
\mbox{e}^{2{\log}f(a) K_{3}}
\right\}
\mbox{e}^{-\frac{c \mbox{e}^{b}}{f(a)} K_{+}}  \nonumber \\
&=&
\mbox{e}^{\frac{c \mbox{e}^{b}}{f(a)} K_{+}}
\left(
\mbox{e}^{2{\log}f(a)}
K_{-}
\right)
\mbox{e}^{-\frac{c \mbox{e}^{b}}{f(a)} K_{+}}  
=
f(a)^{2}
\mbox{e}^{\frac{c \mbox{e}^{b}}{f(a)} K_{+}}
K_{-}
\mbox{e}^{-\frac{c \mbox{e}^{b}}{f(a)} K_{+}}  \nonumber \\
&=&
f(a)^{2}
\left(
K_{-}-2\frac{c \mbox{e}^{b}}{f(a)} K_{3}+
\frac{(c \mbox{e}^{b})^{2}}{f(a)^{2}}K_{3}
\right)    \nonumber \\
&=&
f(a)^{2}K_{-}-2c\mbox{e}^{b}f(a) K_{3}+(c \mbox{e}^{b})^{2}K_{3}. 
\end{eqnarray}
Here we have used the formula
\[
\mbox{e}^{\alpha K_{3}}K_{-}\mbox{e}^{-\alpha K_{3}}=
\mbox{e}^{-\alpha}K_{-},\quad 
\mbox{e}^{\alpha K_{+}}K_{-}\mbox{e}^{-\alpha K_{+}}=
K_{-}-2\alpha K_{3}+{\alpha}^{2}K_{+}. 
\]
Therefore from (\ref{eq:continuation}) and (\ref{eq:continuation-2}) 
we reach 
\begin{equation}
G{'}(a)=K_{-}G(a).
\end{equation}
From the uniqueness of differential equations with the same initial 
condition we obtain $F(a)=G(a)$.

\par \vspace{5mm} 
Similarly we have a formula for some operators based on $su(2)$.
\begin{flushleft}
{\bf Formula}
\end{flushleft}
\begin{equation}
\label{eq:fundamental-formula-su(2)}
\mbox{e}^{aJ_{-}}\mbox{e}^{2bJ_{3}}\mbox{e}^{cJ_{+}}=
\mbox{e}^{xJ_{+}}\mbox{e}^{2yJ_{3}}\mbox{e}^{zJ_{-}}
\end{equation}
where 
\begin{equation}
x=\frac{c\mbox{e}^{b}}{\mbox{e}^{-b}+ac\mbox{e}^{b}},\quad 
y=-{\log}(\mbox{e}^{-b}+ac\mbox{e}^{b}),\quad 
z=\frac{a\mbox{e}^{b}}{\mbox{e}^{-b}+ac\mbox{e}^{b}}.
\end{equation}
We leave the proof to the readers.


\end{document}